\documentclass[aip, reprint, graphicx]{revtex4-1}
\usepackage{graphicx}
\draft 

\begin{document}

\title{Relativistic acceleration of Landau resonant particles as a consequence of Hopf bifurcations.} 



\author{A. Osmane and A. M. Hamza}
\affiliation{Department of Physics, University of New Brunswick, Fredericton, NB, E3B5A3, CANADA}
\email{a.osmane@unb.ca}


\begin{abstract}
Using bifurcation theory on a dynamical system simulating the interaction of a particle with an obliquely propagating wave in relativistic regimes, we demonstrate that uniform acceleration arises as a consequence of Hopf bifurcations of Landau resonant particles. The acceleration process arises as a form of surfatron established through the locking in pitch angle, gyrophase, and physical trapping along the wave-vector direction. Integrating the dynamical system for large amplitudes ($\delta B/B_0\sim0.1$) obliquely propagating waves, we find that electrons with initial energies in the keV range can be accelerated to MeV energies on timescales of the order of milliseconds. The Hopf condition of Landau resonant particles could underlie some of the most efficient energization of particles in space and astrophysical plasmas. 
\end{abstract}


\maketitle 

\section{Introduction}
The main purpose of this Letter is to study the wave-particle interaction in the general case of oblique propagation and in the relativistic limit. The more recent inclusion of oblique waves in modeling is not fortuitous  \cite{Hamza06, Araneda08, Osmane10}. Obliquely propagating waves are not only theoretically predicted \cite{Hollweg02b} but are also observed in numerous and diverse space plasma regimes, e.g., in fast streams of the solar wind \cite{Bavassano82}, upstream of the bow shock \cite{Meziane01}, magnetosheath \cite{Perri09}, cometary environment \cite{Neubauer93}, the Earth's plasma sheet boundary \cite{Broughton08} and the Van Allen radiation belt \cite{Catell08}. The interest in obliquely propagating waves mainly resides in the inclusion of an electric field along the background magnetic field. Unlike the parallel propagation case and due to constraints imposed by Maxwell-Lorentz invariant quantities, it is impossible to find an inertial frame of reference for which the electric field vanishes\cite{Jackson}. The electric field of the wave can therefore provide acceleration mechanisms and/or physical trapping along the background magnetic field. Similarly, the study of relativistic regimes to modeling space and astrophysical plasmas now appears more than ever as absolute necessity, and more so for the numerous problems (i.e. cosmic rays, radiation belt electrons) where the acceleration or/and injection of charged particles to relativistic energies remains  misunderstood. In the following report, we ignore the more commonly studied case of cyclotron-resonant particles ($\omega \sim\Omega$) and concentrate instead on the parameter space where the Landau resonance ($\omega \ll \Omega$) is accessible. Even though cyclotron-resonant interactions are observed and understood to play an important role in the kinetic description of space and astrophysical plasmas, other means of energy exchange between waves and particles have been lurked behind a predominant focus on cyclotron-resonance. For instance, the perpendicular heating in fast streams of the solar wind could as well result from trapped particles caught in broadened Landau resonance instead of the commonly assumed cyclotron resonance \cite{Lehe09, Osmane10}. This report also aims at providing further ground for such views, but in the context of relativistic and weakly collisional plasmas. \\
\section{Dynamical System}
This problem is addressed by using dynamical systems' theory, which although lacking levels of self-consistency that simulations provide, can facilitate the understanding of complex systems such as plasmas and provide for an intuitive leap between theoretical models and simulations. 
Hence, our study begins by writing the equation of motion for a particle in an electromagnetic field as follow : \begin{equation}
\frac{d\textbf{p}}{dt}= e\bigg{[}\textbf{E}(\textbf{x},t)+\frac{\textbf{p}}{m\gamma c} \times \textbf{B}(\textbf{x},t)\bigg{]}
\end{equation}
for a particle of momentum $\textbf{p}=m\gamma\textbf{v}$, rest mass $m$, charge $e$ and Lorentz contraction factor $\gamma=\sqrt{1+p^2/m^2c^2}$.  The fields topology consist of an electromagnetic obliquely propagating wave of amplitude $(\delta \mathbf{E}, \delta \mathbf{B})$ superposed on a  background magnetic field $\mathbf{B}_0$.
We choose the electromagnetic wave vector $\mathbf{k}$ to point in the $\hat{z}$ direction and the background magnetic field to lie in the $y-z$ plane. Hence, the propagation angle, $\theta$, denoting the obliqueness of the wave, is defined as $\mathbf{k} \cdot \mathbf{B}_0=kB_0\cos(\theta)$. The magnetic field components of the wave are written as 
\begin{equation}
\left\{ 
\begin{array}{l l} 
\delta B_x= \delta B \sin(kz-\omega t)\\ 
\delta B_y=\delta B \cos(kz-\omega t),\\   
\end{array} \right.
\end{equation}
with the electric components provided by Faraday's law, $c\mathbf{k} \times \delta \mathbf{E} (\mathbf{k},\omega) =\omega \delta \mathbf{B}(\mathbf{k},\omega)$.
We can then write the dynamical system equations for the chosen electromagnetic field topology in terms of the following variables : $v_\Phi=\omega/k$, $p_\Phi=m\gamma v_\Phi$, $\Omega_1=e\delta B/mc\gamma$, $\Omega_0=e B_0/mc\gamma$. Hence, obtaining the following set of equations :  
\begin{equation}
\left\{ 
\begin{array}{l l} 
\dot{p}_x=p_y\Omega_0\cos(\theta)+(p_\Phi-p_z)\Omega_1\cos(kz-\omega t)  +p_z\Omega_0\sin(\theta)\\
\dot{p}_y=-p_x\Omega_0\cos(\theta)+(p_z-p_\Phi)\Omega_1\sin(kz-\omega t)\\
\dot{p}_z=-p_x\Omega_0 \sin(\theta)+p_x\Omega_1\cos(kz-\omega t) -p_y\Omega_1\sin(kz-\omega t)\\
\dot{z}=p_z v_\Phi/p_\Phi
\end{array} 
\right.
\end{equation} 
It follows that the dynamical gyrofrequencies $(\Omega_0, \Omega_1)$ can be tracked as the fifth variable of the dynamical system : 
\begin{equation}
\dot{\Omega}_0=\frac{d}{dt}\bigg{(}\frac{eB_0}{mc\gamma}\bigg{)} \nonumber =-\Omega_0 \frac{pc^2}{m^2c^4+p^2c^2}\dot{p}.
\end{equation}
We now proceed by eliminating the time dependence  by making the following change of variables : 
\begin{equation}
p_x'=p_x, \hspace{.5mm}p_y'=p_y, \hspace{.5mm} p_z'=(p_z-p_\phi), \hspace{.5mm} z'=(z-v_\phi t).
\end{equation}
Hence, we write the the dynamical system in terms of the primed variables as follow : \begin{equation}
\label{eq:ds_in_ps}
\left\{ 
\begin{array}{l l} 
\dot{p}_x'=\Omega_0 p_y'\cos(\theta)-\Omega_1p_z' \cos(kz')  +\Omega_0 (p_z'+p_\phi) \sin(\theta)\\
\dot{p}_y'=-\Omega_0p_x'\cos(\theta)+\Omega_1p_z' \sin(kz')\\
\dot{p}_z'=-\Omega_0p_x'\sin(\theta)+\Omega_1(\frac{n^2-1}{n^2})(p_x'\cos(kz') -p_y'\sin(kz'))\\
\dot{z}'=p_z'v_\Phi/p_\Phi\\
\end{array} 
\right.
\end{equation} 
with the refractive index $n^2=c^2/v_\Phi^2$. In addition to equation $(4)$, one can compute the particle orbits for a class of parameters $\theta, n, \delta_1=\Omega_0/\Omega_1$ and $\delta_2=\omega/\Omega_0\gamma$.\\
\section{Hopf bifurcations of Landau resonant trapped orbits---} Despite the apparent simplicity of the dynamical system, the inclusion of the relativistic terms results in a number of interesting properties that extend beyond the scope of this report. We hereafter focus on one of these properties arising from the bifurcation in stability of fixed (stationary) points. Indeed, it can easily be shown that the set of equations $(6)$ possesses a class of fixed (stationary) points that can be represented as follow : 
\begin{eqnarray}
\label{eq:FP5}
p_{x0}'=p_{z0}'=0; \hspace{8mm} p_{y0}' = -p_\Phi\tan(\theta)\nonumber \\
\gamma_0=\frac{1}{\sqrt{1-\frac{v_\Phi^2}{c^2}(1+\tan^2(\theta))}}; \hspace{8mm} Z=kz'=0,\pi. \nonumber
\end{eqnarray}
\begin{figure}[tb] 
   \centering
   \includegraphics[width=0.45\textwidth]{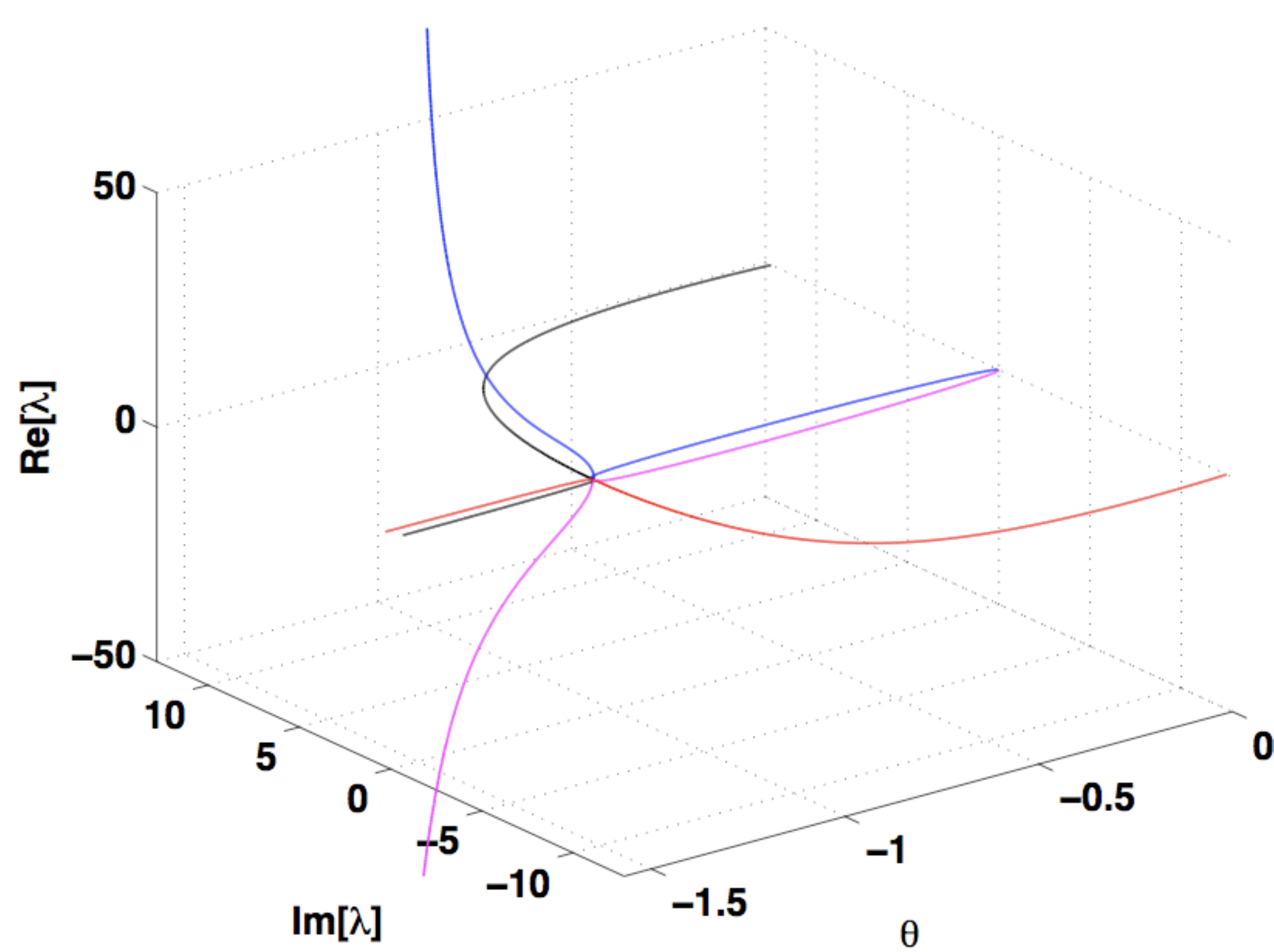}
   \caption{Eigenvalues' dependence on the propagation angle $\theta$ for fixed parameters $\delta_1=0.1$, $\delta_2=0.0696$, $n^2=2$ and the fixed point of component $Z_0=0$. The bifurcation through the positive real axis takes place for propagation angle $\theta_c=60^o$. The fixed point is stable for $\theta < \theta_c$ and unstable for $\theta>\theta_c$.}
   \label{fig:example}
\end{figure}
These points are informative of the values for which a particle is physically trapped by the electromagnetic field. Their values in velocity space correspond to the Landau resonance condition. This can be more clearly seen if one applies the inverse of the translation in equation (5) followed by a rotation in a system of coordinate with the $z$ axis parallel to the background magnetic field\cite{Osmane10}. The next fundamental step in dynamical system theory is to investigate the stability of the fixed points. In order to do so, we apply a basic Lyapunov linear analysis that can be found in any textbook on dynamical systems\cite{Regev}. 
Hence, solving the eigenvalue problem $(\mathbf{J}-\lambda\mathbf{I})=0$ for the Jacobian $\mathbf{J}$ and eigenvalue $\lambda$, we find a bi-quadratic polynomial function in $\lambda$ that can be written as $\chi(\lambda)=\lambda^4+\eta_1\lambda^2+\eta_2=0$, with the constant coefficients $\eta_1$ and $\eta_2$ given by the following expressions : 
\begin{eqnarray}
\eta_1&=&\frac{\delta_1}{\delta_2\gamma_0}\frac{n^2-1}{n^2}\tan(\theta)+\frac{\cos^2(\theta)}{\delta_2^2\gamma_0^2}\nonumber\\
  &-&\frac{\delta_1}{\delta_2^2\gamma_0^2}\bigg{(}-\frac{n^2-1}{n^2}\pm2\sin(\theta)-\frac{\sin^2(\theta)}{\delta_1}\mp\frac{\sin(\theta)}{n^2}\bigg{)}\nonumber\\
 \eta_2&=&\frac{\delta_1}{\delta_2^2\gamma_0^2}\frac{n^2-1}{n^2}\sin(\theta)\cos(\theta),
\end{eqnarray}
 \begin{figure}[tb] 
   \centering
   \includegraphics[width=0.45\textwidth]{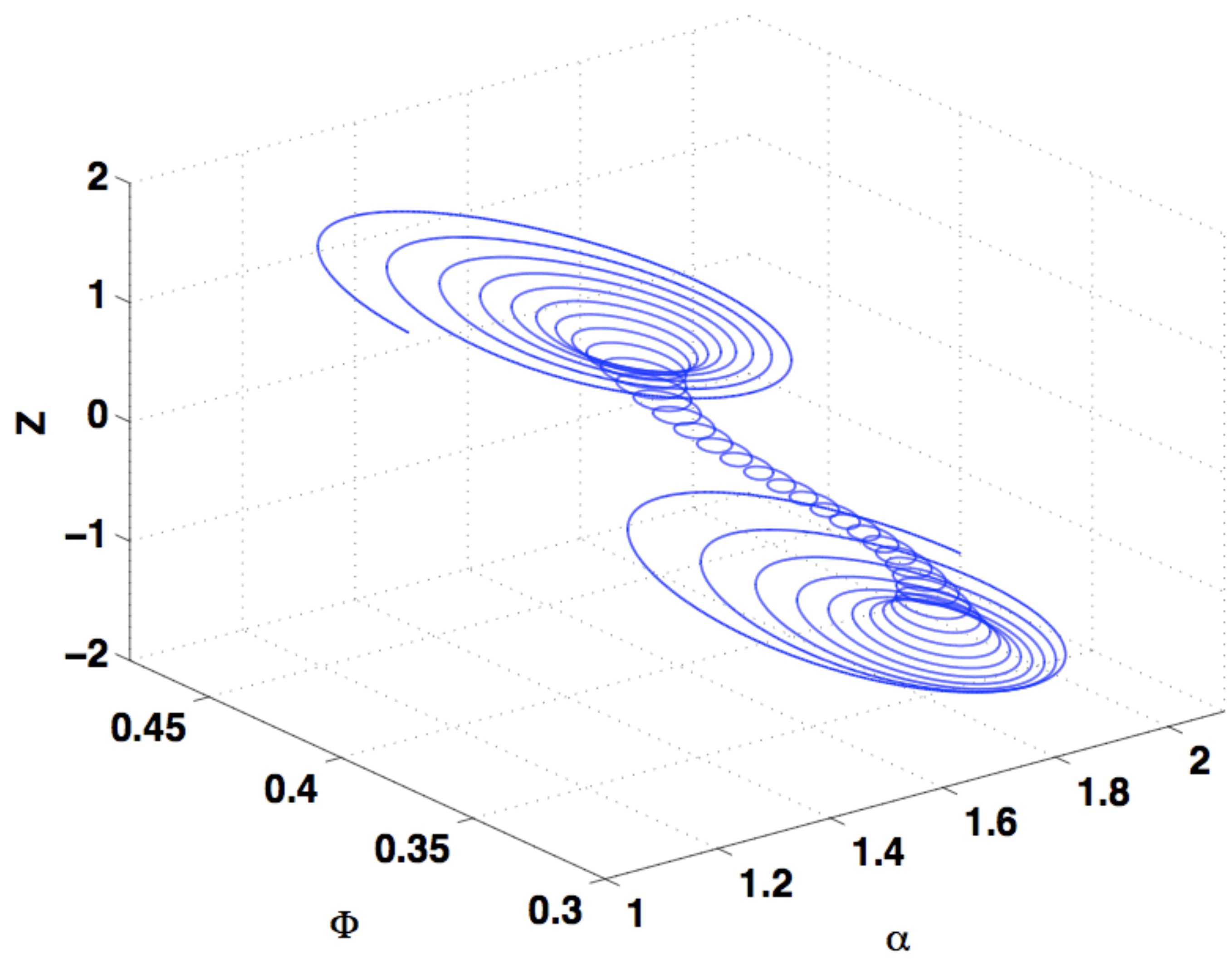}
   \caption{Particle orbits for parameters $\delta_1=0.1$, $\delta_2=0.0696$, $n^2=4$, $\theta=\theta_c-1^o$ and initial conditions $v_{x0}'=0$, $v_{y0}'=-v_\Phi\tan(\theta)-1.6v_\Phi$, $v_{z0}'=-v_\Phi$, $Z_0'=0$.}
   \label{fig:example3}
\end{figure}
with the $\pm$ symbol denoting the values for $Z=0$ and $Z=\pi$. A close look at the coefficients of equation set $(7)$ shows that all four eigenvalues will cross the zero real axis when the condition
 \begin{equation}
 n^2-1=\tan^2(\theta)
 \end{equation}
 is respected. That is, for parameter values corresponding to $\gamma_0^{-1}=0$ and resulting in $\lambda^4=0$. This condition can be more clearly expressed through Figure 1 where we plotted the dependence of the real and imaginary part of all four eigenvalues as a function of $\theta$, while keeping the remaining parameters ($\delta_1, \delta_2, n^2$) constant.  We observe that when the condition in equation(8) is respected, the equilibrium evolves from stable to unstable equilibrium since the real part of one of the eigenvalues becomes positive. This type of bifurcation, where pairs of complex conjugate eigenvalues cross through the imaginary axis, is the well-known Hopf bifurcation \cite{Regev}. The fixed point for $Z=\pi$ is linearly unstable for every parameter range and values chosen in Figure 1 with the exception of the parameters for which equation (8) is respected. The eigenvalue profile and the following conclusions do not differ significantly for low-frequencies ($\delta_2 < 1$) and large-amplitudes ($\delta_1 \leq 1$), that is the relevant range of parameters in space and astrophysical plasmas.\\ 
\begin{figure}[tb] 
   \centering
   \includegraphics[width=0.45\textwidth]{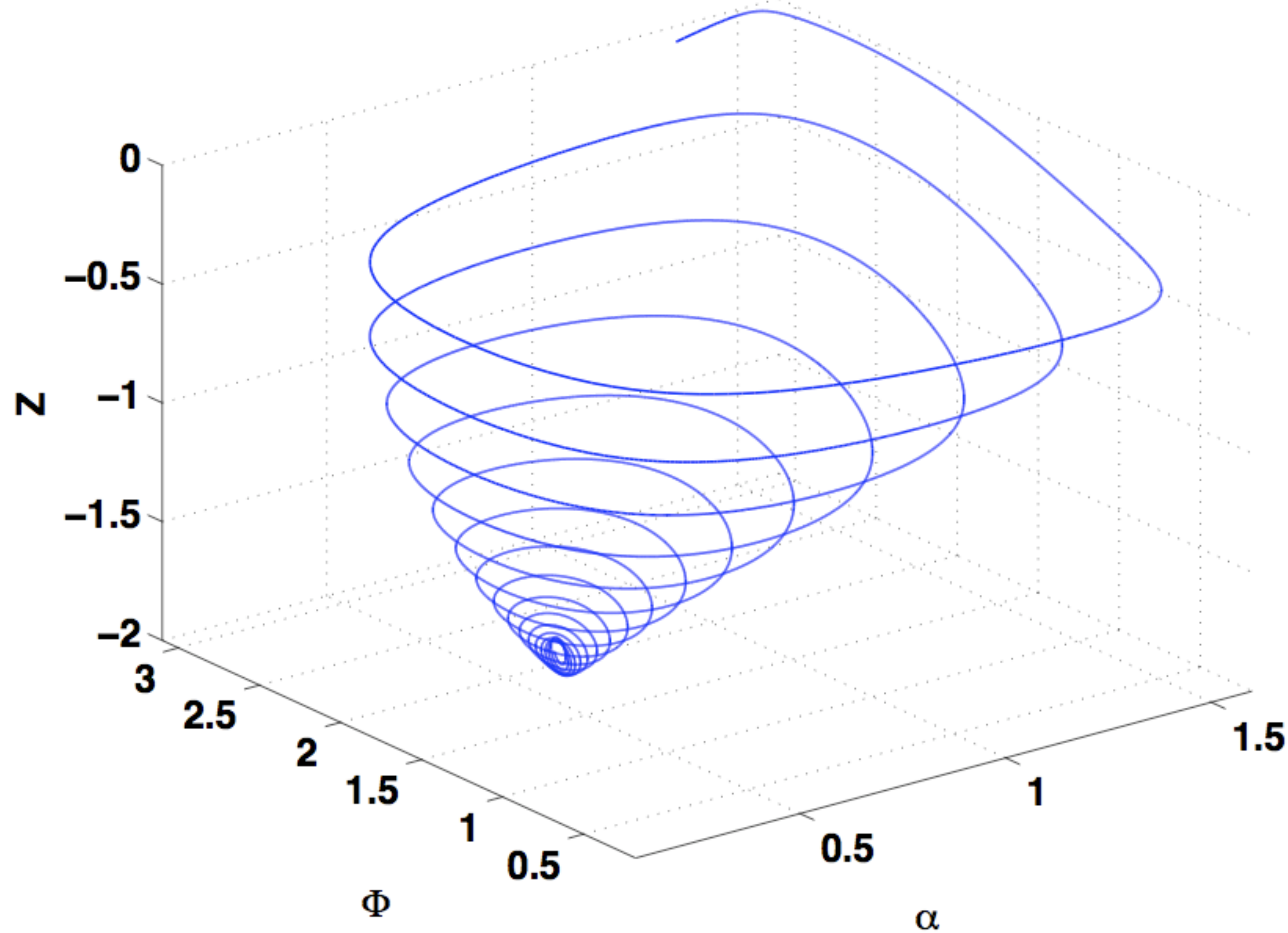}
   \caption{Particle orbits for parameters $\delta_1=0.1$, $\delta_2=0.0696$, $n^2=4$, $\theta=\theta_c$. The orbit is locked in phase-space and trapped along $Z$.}
   \label{fig:example1}
\end{figure}
\begin{figure}[tb] 
   \centering
   \includegraphics[width=0.45\textwidth]{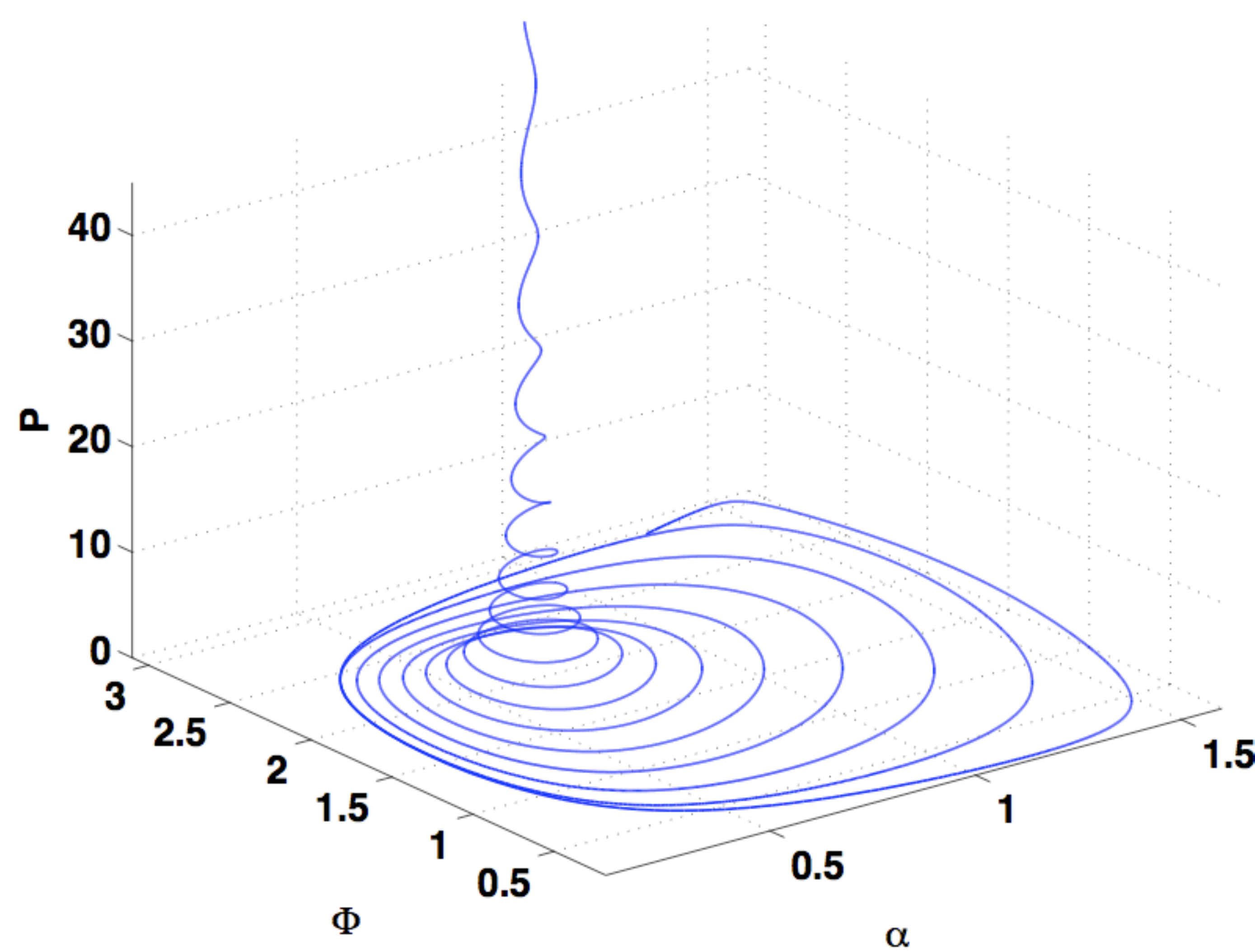}
   \caption{Particle orbits for parameters $\delta_1=0.1$, $\delta_2=0.0696$, $n^2=4$, $\theta=\theta_c$. The orbit is uniformly accelerated once locked in phase-space.}
   \label{fig:example2}
\end{figure}
\begin{figure}[tb] 
   \centering
   \includegraphics[width=0.45\textwidth]{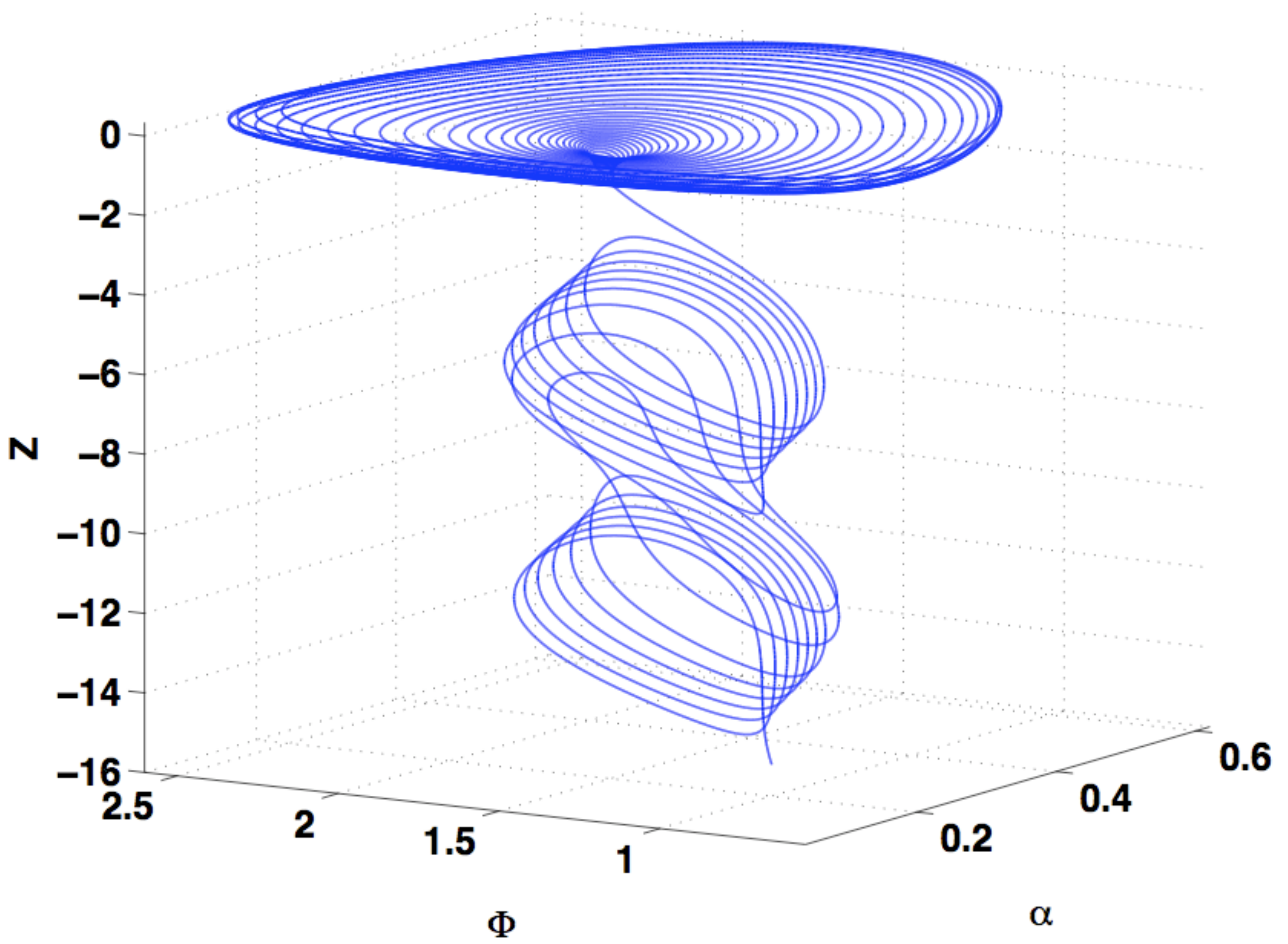}
   \caption{Particle orbits for $\theta>\theta_c$. A torus emerge from the two Hopf bifurcations.}
   \label{fig:example4}
\end{figure}
\section{Numerical integration---} We now investigate the effects of the bifurcation on the particles belonging to the basin of attraction of the fixed point with component $Z=0$. For the sake of clarity and in order to underlie the physical processes at play, we choose to represent the particle momentum in terms of spherical coordinates $(p',\alpha, \Phi)$ instead of the cartesian coordinates $(p_x', p_y', p_z')$ in equation set $(6)$. The transformation from one expression of the momentum through the other can be made using the following definitions for the magnitude $p'=\sqrt{p^{'2}_x+p^{'2}_y+p^{'2}_z}$, the pitch angle $\tan (\alpha) = \frac{p'_\perp}{p'_{\parallel}}$, and the dynamical gyrophase $\tan (\Phi) = \frac{p'_{\perp 1}}{p'_{\perp 2}}=\frac{p_x'}{p_y'\cos(\theta)+p_z'\sin(\theta)}$, where the parallel and perpendicular symbols are with respect to $\mathbf{B_0}$. In Figure 2 the orbit of a particle interacting with a large amplitude and low-frequency wave is shown for $\alpha$, $\Phi$ and $Z$ for a propagation angle nearly obeying the condition of equation $(8)$. For such parameter values, the fixed point is linearly stable and a particle belonging to the basin of attraction will remain physically trapped along the wave-vector. The orbit eventually closes onto itself as the particle bounces back and forth in the wave potential, alternatively loosing and gaining energy with no net gain over one period. If we modify the propagation angle such that the condition in equation $(8)$ is respected, we can see from Figures 3 and 4 that a particle will asymptotically converge into a point in the $(\alpha, \Phi)$ phase-space while it gets trapped along $Z$ and subsequently diverges to infinity in momentum. That is, the fixed point becomes an attractor along $(\alpha, \Phi)$, and the particles initially belonging to the basin of attraction will be locked forever in phase-space and experience uniform acceleration through the constant electric field observed by the particle. If we increase the propagation angle further, such that $\theta>\theta_c$, we find that two dimensional torus are created (see Figure 5) and that particles can be neither physically trapped nor uniformly accelerated. However, even though the dynamics for $\theta>\theta_c$ deserves a study of its own, we now focus on the case $\theta\sim\theta_c$ for which particles can be energized irreversibly.\\
\begin{figure}[tb] 
   \centering
   \includegraphics[width=0.40\textwidth]{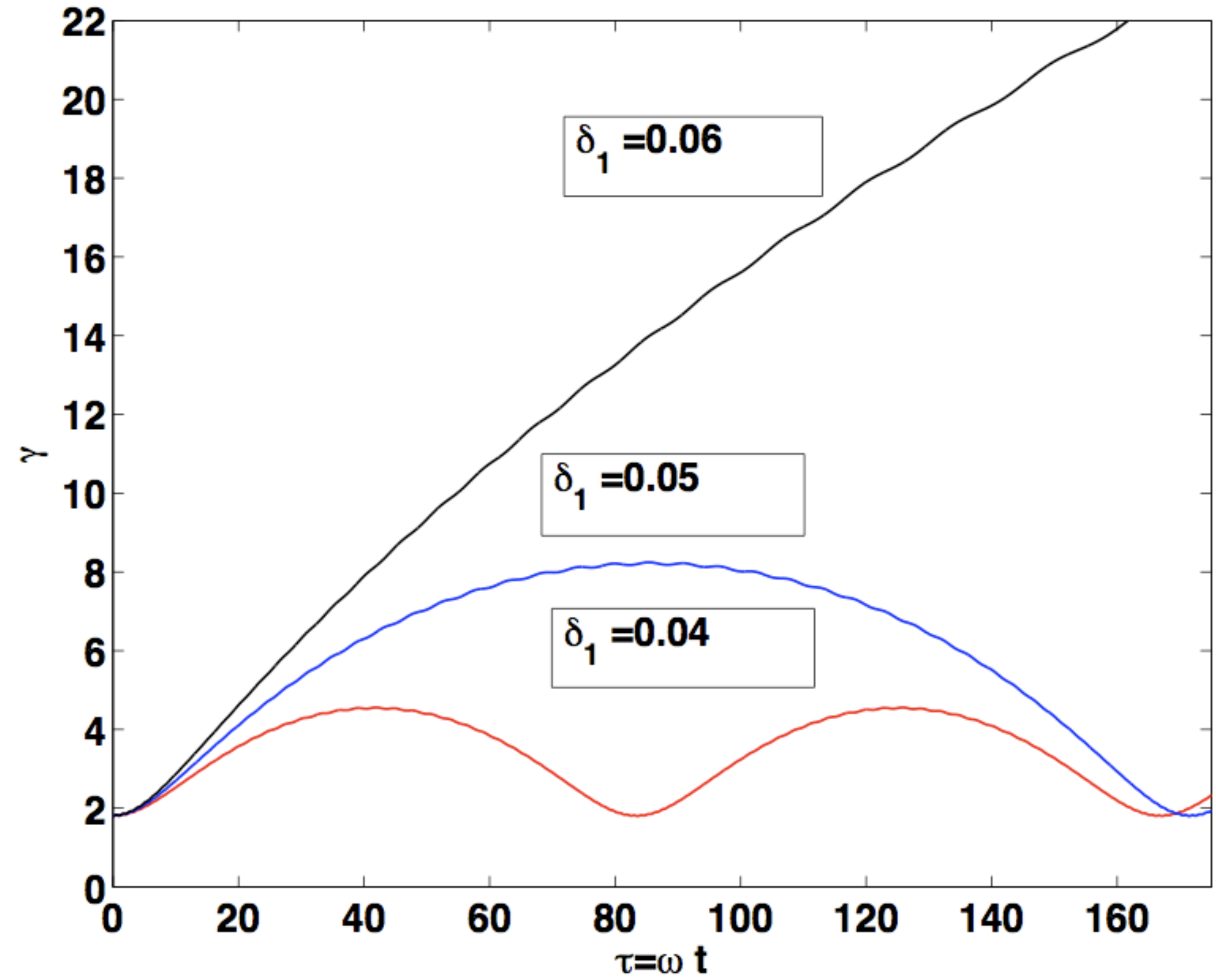}
   \caption{Lorentz factor $\gamma$ as a function of time for $\delta_1=(0.04, 0.05, 0.06)$, $\delta_2=0.1$, $n^2=9$, $\theta=\theta_c$ and initial condition $v_x=v_z\sim 0, v_y=-v_\Phi\tan(\theta_c)-v_\Phi/3, Z=kz'=0$. The particle can be accelerated to MeV energies for time scales of less than a millisecond.}
   \label{fig:example4}
\end{figure}
\section{Discussion---} Phase-locking and trapping of a particle while submitted to a constant electric field has been predicted and observed before in previous models using Hamiltonian and/or asymptotic approaches. Indeed, similar processes for a wide range of electrostatic and/or electromagnetic topologies have been referred to as {\it{surfatron}} \cite{{Katsouleas, Karimabadi90, Chernikov92}}. The novel result reported here is that the mechanism arises as a result of Hopf bifurcations at the Landau resonance. Qualitatively, one could describe this result by the statement that there are {\it{resonances of Landau resonances}} resulting in efficient energization. \\
In order for this mechanism to be reproducible in space and astrophysical plasmas, the acceleration should take place for realistic wave amplitudes, and belong to a sufficiently wide volume of velocity-space to affect a portion of a distribution functions. In other words, the basin of attraction should be wide enough, and particles of moderate energies should be able to be attracted into it. 
Choosing a highly oblique wave $\theta\sim 71^o$, with a wave frequency given by $\delta_2=0.1$ and amplitudes of the order of a percent of the background magnetic field, we integrate the dynamical system for a few wave periods. The results are shown in Figure 6 for different wave amplitudes.  Whereas the irreversible acceleration, coinciding with the Hopf bifurcations in phase space, requires a wave amplitude of $\delta_1\geq 0.06$ capable of physical trapping along $Z$; particles can still be brought to relativistic energies if they are caught in sufficiently close to the fixed point. The size of the basin of attraction varies significantly as a function of the parameter $\delta_2$ but we nonetheless find that particles with moderate energies ranging from few keV to hundreds of keV, can be accelerated to relativistic levels on timescales comparable to the gyroperiod. The basin of attraction is therefore wide enough to affect a significant portion of a distribution function. In the case shown in Figure 6 for $\delta_2=0.1$, electrons with few hundreds keV are accelerated to MeV energies on timescales less than a millisecond. Such timescales of energization for large-amplitudes $\delta_1\sim 0.06$ and low frequency $\delta_2\sim 0.1$ suggest that this mechanism could be of interest for studies of space (e.g. radiation belts) and cosmic (e.g. galactic plasma) plasmas if permeated by large-amplitude oblique waves. Further work is now currently underway to apply methods presented in this paper to the specific problem of electron acceleration in the planetary radiation belts.  

\begin{acknowledgments}
 We thank Dr. K. Meziane for helpful discussions. This work was supported by the Natural Sciences and Engineering Research Council of Canada (NSERC). 
 \end{acknowledgments}

\end{document}